\newcommand{\Rb}{$^{87}$Rb}
\newcommand{\imb}{\ensuremath{\Theta}}
\newcommand{\imbbar}{\ensuremath{\overline{\Theta}}}
\newcommand{\Ul}{\ensuremath{U_{\textsf{l}}}}
\newcommand{\Us}{\ensuremath{U_{\textsf{s}}}}
\newcommand{\UlUs}{\Ul{}/\Us}
\newcommand{\Vx}{\ensuremath{V_{\textsf{x}}}}
\newcommand{\Vy}{\ensuremath{V_{\textsf{y}}}}
\newcommand{\Vz}{\ensuremath{V_{\textsf{z}}}}
\newcommand{\Dc}[1]{\ensuremath{\Delta_{c}^{\textsf{#1}}}}
\newcommand{\dd}[1]{\ensuremath{\delta_{\textsf{#1}}}}
\newcommand{\ddtp}[1]{\ensuremath{\delta_{\textsf{#1}}/2\pi}}

\newcommand{\Dctp}[1]{\ensuremath{\Delta_{\textsf{c}}^{\textsf{#1}}/2\pi}}
\newcommand{\Er}[1]{\ensuremath{E_{\textsf{R}}^{\textsf{#1}}}}

\newcommand{\Ham}{\ensuremath{\hat{\mathcal{H}}}}
\newcommand{\dE}[1]{\ensuremath{\Delta E_{\text{#1}}}}
\newcommand{\del}[1]{\ensuremath{\delta_{\text{#1}}}}
\newcommand{\taus}[1]{\ensuremath{\tau_{\text{s}}^{#1}}}
\newcommand{\nph}{\ensuremath{n_{\textsf{ph}}}}

\pdfoutput=1

\documentclass[aps,prl,9pt,twocolumn,superscriptaddress,a4paper,floatfix,nobibnotes]{revtex4}
\usepackage{color}
\usepackage{sfmath}
\usepackage{graphicx}
\usepackage{amsmath,amssymb}
\usepackage{hyperref}
\usepackage{float}

\begin{document}
\title{Metastability and avalanche dynamics in strongly-correlated gases with long-range interactions}
\author{Lorenz Hruby} 
\affiliation{Institute for Quantum Electronics, ETH Zurich, 8093 Zurich, Switzerland}
\author{Nishant Dogra}  
\affiliation{Institute for Quantum Electronics, ETH Zurich, 8093 Zurich, Switzerland}
\author{Manuele Landini}  
\affiliation{Institute for Quantum Electronics, ETH Zurich, 8093 Zurich, Switzerland}
\author{Tobias Donner} 
\email{donner@phys.ethz.ch}
\noaffiliation
\affiliation{Institute for Quantum Electronics, ETH Zurich, 8093 Zurich, Switzerland}
\author{Tilman Esslinger} 
\affiliation{Institute for Quantum Electronics, ETH Zurich, 8093 Zurich, Switzerland}
\date{\today}

\begin{abstract}
We experimentally study the stability of a bosonic Mott-insulator against the formation of a density wave induced by long-range interactions, and characterize the intrinsic dynamics between these two states. The Mott-insulator is created in a quantum degenerate gas of 87-Rubidium atoms, trapped in a three-dimensional optical lattice. The gas is located inside and globally coupled to an optical cavity. This causes interactions of global range, mediated by photons dispersively scattered between a transverse lattice and the cavity. The scattering comes with an atomic density modulation, which is measured by the photon flux leaking from the cavity. We initialize the system in a Mott-insulating state and then rapidly increase the global coupling strength. We observe that the system falls into either of two distinct final states. One is characterized by a low photon flux, signaling a Mott insulator, and the other is characterized by a high photon flux, which we associate with a density wave. Ramping the global coupling slowly, we observe a hysteresis loop between the two states – a further signature of metastability. A comparison with a theoretical model confirms that the metastability originates in the competition between short- and global-range interactions. From the increasing photon flux monitored during the switching process, we find that several thousand atoms tunnel to a neighboring site on the time scale of the single particle dynamics. We argue that a density modulation, initially forming in the compressible surface of the trapped gas, triggers an avalanche tunneling process in the Mott-insulating region. 
\end{abstract}

\maketitle

When found in a metastable state or phase, a system resides in a condition differing from its state of least energy for an extended period of time. Examples for long-lived metastable phases are found in magnetized materials, glasses, crystals like diamond, as well as in macromolecules \cite{Anderson1972, Karplus2002, Brazhkin2006}. In many solid-state systems, metastability can be described by a first-order phase transition \cite{Binder1987}, yet the less accessible switching dynamics and its associated time scales are crucial to gain insights into the mechanisms of structure formation.

\begin{figure}[!h]
\centering
\includegraphics[width=0.99 \columnwidth]{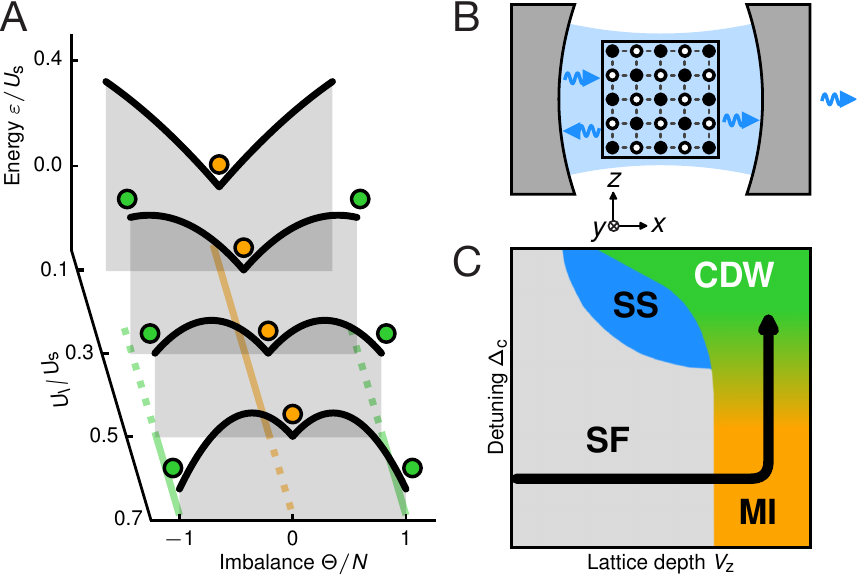}
\caption{Metastability and system overview. (A) Mean-field results from the toy model. In the presence of short-range interactions \Us{} and global-range interactions \Ul{} atoms placed in a lattice potential can show metastable behavior. States (indicated by circles) can be protected by an energy barrier and the present state of the system depends on its history, leading to hysteresis. The Mott insulator (orange line) and the charge density wave (green lines) are either stable (solid), metastable (dashed) or unstable. (B) Our system consists of a Bose-Einstein condensate coupled to a single mode of an optical resonator in the presence of 3D optical lattices. The atoms can create a particle imbalance \imb{} by arranging in a checkerboard pattern which maximizes scattering of photons from a $z$ lattice (not shown) into the resonator mode. (C) Schematic phase diagram of the system with a superfluid (SF), a lattice supersolid (SS), a Mott insulator (MI) and a charge density wave (CDW) phase. The black arrow illustrates the experimental sequence: We prepare the atoms in the SF phase and ramp up the 3D optical lattices to increase \Us{}, which brings the system into the MI phase. Subsequently, we carry out a detuning ramp towards cavity resonance which increases \Ul{}.}
\vspace{-3em}
\label{fig:system}
\end{figure}

Ultracold atoms emerge as a promising tool to study questions related to metastability in quantum many-body systems, due to the precise knowledge and high-level of control over the underlying Hamiltonian. Indeed, metastable states, many-body localization, and first-order phase transitions have recently attracted theoretical \cite{Menotti2007, Gopalakrishnan2011, Strack2011, Altman2015, Andraschko2014, Eisert2015} and experimental interest \cite{Haller2009, Eckel2014, Schreiber2015, Campbell2016, Kadau2016, Trenkwalder2016}. The presence of long-range interactions is of particular importance to induce and influence metastability, since it makes decay processes like nucleation and phase separation energetically costly, resulting in increased lifetimes of higher energy states, as recently observed in Rydberg excitation clusters \cite{Letscher2016}. The consequences are even more severe in systems with long-range interactions decaying slower than $1/r^d$, where $r$ is the inter-particle distance and $d$ is the dimensionality of the system, as a separation into independent clusters is no longer possible. The lifetime of metastable phases then scales with the system size and diverges in the thermodynamic limit \cite{Antoni1995, Mukamel2005}.

In our experiment the global interactions arise from the coupling of a Bose-Einstein condensate (BEC) to a single mode of an optical high-finesse cavity \cite{Baumann2010, Mottl2012}. With the atomic gas trapped in a three-dimensional (3D) optical lattice we can simultaneously control short-range interactions and push the system into a strongly correlated regime (Fig.~\ref{fig:system}B). The phase diagram of the system is schematically shown in Fig.~\ref{fig:system}C. It was recently determined experimentally \cite{Klinder2015insulator, Landig2016} and studied theoretically \cite{Li2013, Bakhtiari2015, Caballero-Benitez2015, Chen2016, Dogra2016, Niederle2016, Sundar2016, Panas2017, Flottat2017}. In the thermodynamic limit a first order phase transition from a Mott insulator (MI) \cite{Jaksch1998, Greiner2002} to a charge density wave (CDW) state has been predicted \cite{Chen2016, Dogra2016, Sundar2016, Flottat2017}. 

\section*{Toy model}
To achieve a basic understanding of our system we study a toy model with Hamiltonian $\Ham{} = \frac{1}{2}\Us{} \sum_{i\in e,o} \hat{n}_{i}\left(\hat{n}_{i}-1\right) - \frac{1}{K}\Ul{} \hat{\Theta}^2$, i.e. an extended Bose-Hubbard model where we have neglected tunneling for simplicity. We consider the situation of a fixed number of atoms $N$ in a box potential, with $K=N$ lattice sites and an average filling per lattice site of $\left<\hat{n}_i\right>=1$. $\Us{}$ and $\Ul{}$ denote the strength of short- and global-range interactions, respectively. The global-range interaction term favors a particle imbalance between even and odd lattice sites. It is characterized by the imbalance operator $\hat{\Theta} = \sum_{i\in e}\hat{n}_i - \sum_{i\in o}\hat{n}_i$, where $\hat{n}_i$ counts the number of atoms on lattice site $i$ and the sub-indices $e$ and $o$ denote even and odd lattice sites, respectively.

We obtain the average ground state energy per particle $\varepsilon=\left<\Ham{}\right>/N$ as a function of the imbalance $\imb{}=\left<\hat{\Theta}\right>$ for varying \UlUs{}, see Fig.~\ref{fig:system}A (SI Appendix). When global-range interactions are weak ($\UlUs{} < 0.25$), the free energy landscape has a single global minimum at imbalance $\imb{}=0$ corresponding to a Mott insulator (MI) with exactly one atom on every lattice site. For $\UlUs{} > 0.5$ global-range interactions dominate and we find an insulating ground state with a modulated density distribution which we denote charge-density wave (CDW). Since the discrete even-odd symmetry of the lattice is broken, the energy landscape shows two global minima at $\imb{}/N=\pm 1$. In the region around $\UlUs{}\approx 0.5$ this model shows metastable behavior \cite{Panas2017, Flottat2017}. Here the MI state is a local minimum in the free energy landscape, separated from the CDW states by an energy barrier, which results from the competition between strong interactions of short- and global-range character. 

\begin{figure}[t]
\centering
\includegraphics[width=\columnwidth]{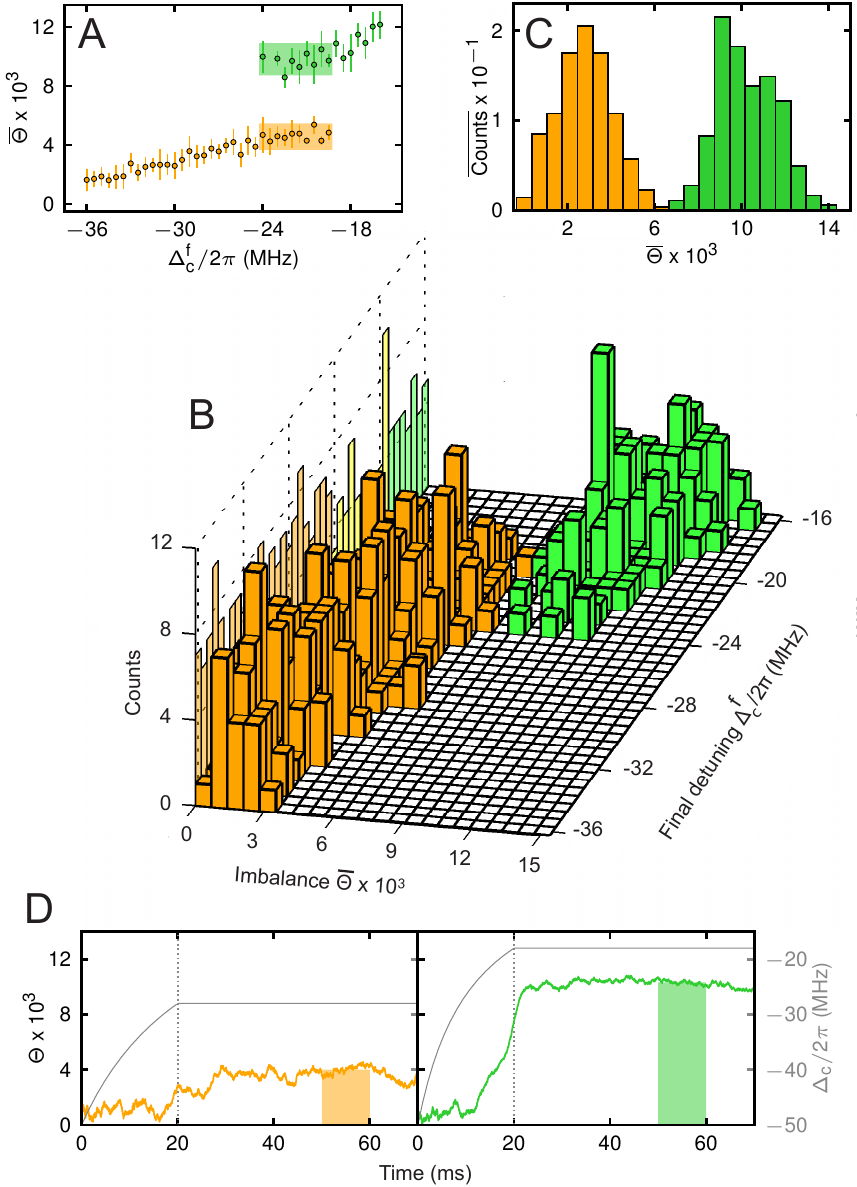}
\caption{\textit{Metastability Measurement.} Observation of two distinct steady-state imbalances \imbbar{}, shown in orange and green, and exemplary time-traces. We prepare an MI and then quench the detuning from $\Dctp{}=-50$~MHz to \Dc{f} closer to resonance within $20$~ms, increasing \Ul{}. (A) Mean values of the imbalance \imbbar{}, errors are SD. The imbalance \imbbar{} is separated by a gap of $5.2(1.4) \times 10^3$ atoms into two levels. (B) Histogram as a function of \imbbar{} and \Dc{f} with bin sizes of $700$~atoms in \imbbar{} and $0.5$~MHz in \Dc{f}. The left plane depicts the maximum number of counts observed at every \Dc{f}. (C) Histogram of the normalized sum of all counts with respect to \imbbar{}, for the normalization see SI Appendix. (D) Exemplary time-traces for quenches ending at $\Dctp{f}=-28$~MHz (left) and $\Dctp{f}=-18$~MHz (right). The shaded regions indicate where the averaged imbalance \imbbar{} is extracted. This experiment was performed with $25(2)\times 10^3$~atoms at maximum lattice depths of $(\Vx{}, \Vy{}, \Vz{})=(17.3~\Er{785}, 30.7~\Er{671}, 11.1~\Er{785})$.}
\label{fig:metastable}
\end{figure}
\section*{System description}
We load a BEC of $(15 - 25)\times10^3$~\Rb{} atoms into a harmonic potential centered at the position of the cavity mode. The cloud is split into about $70$~weakly coupled two-dimensional~(2D) layers using an optical lattice of $(26.2-30.7)$~\Er{671} depth along the $y$ axis at wavelength $\lambda_{\textsf{y}} = 671.0$~nm (SI Appendix). We specify lattice depths in units of the recoil energy ${\Er{$\lambda$} = h^2/(2m\lambda^2)}$ for the wavelength $\lambda$, where $h$ denotes Planck's constant and $m$ is the atomic mass of \Rb{}. The 2D layers are exposed to a square lattice composed of a free space lattice in the $z$ direction and an intra-cavity optical standing wave along the $x$ direction which is externally applied through the cavity mirrors (Fig.~\ref{fig:system}B) at wavelengths $\lambda_{\textsf{x}} = \lambda_{\textsf{z}} = 784.7$~nm. In all experiments, the depths of these lattices are tuned simultaneously such that $\Vx{}\approx \Vz{}$ (SI Appendix), but due to the special role of the $z$ lattice we refer to \Vz{} throughout the paper. The $z$ lattice mediates global-range atom-atom interactions of tunable strength $\Ul{}\propto \Vz{} / \Dc{}$ via off-resonant scattering into the optical resonator mode \cite{Mottl2012} (SI Appendix). Here \Dc{} is the detuning of the frequency of the laser forming the $z$ lattice from cavity resonance. We estimate a final filling of at most two atoms per lattice site at the center of the cloud in the MI phase. We monitor in real-time the flux of photons leaking out of the cavity using a heterodyne detector \cite{Landig2015}. The flux is converted into an imbalance $\imb{}\propto \sqrt{n_{\textsf{ph}}}$, where $n_{\textsf{ph}}$ represents the mean intra-cavity photon number. For further information on the system see \cite{Landig2016} and SI Appendix.

\begin{figure}[tb]
\centering
\includegraphics[width=\columnwidth]{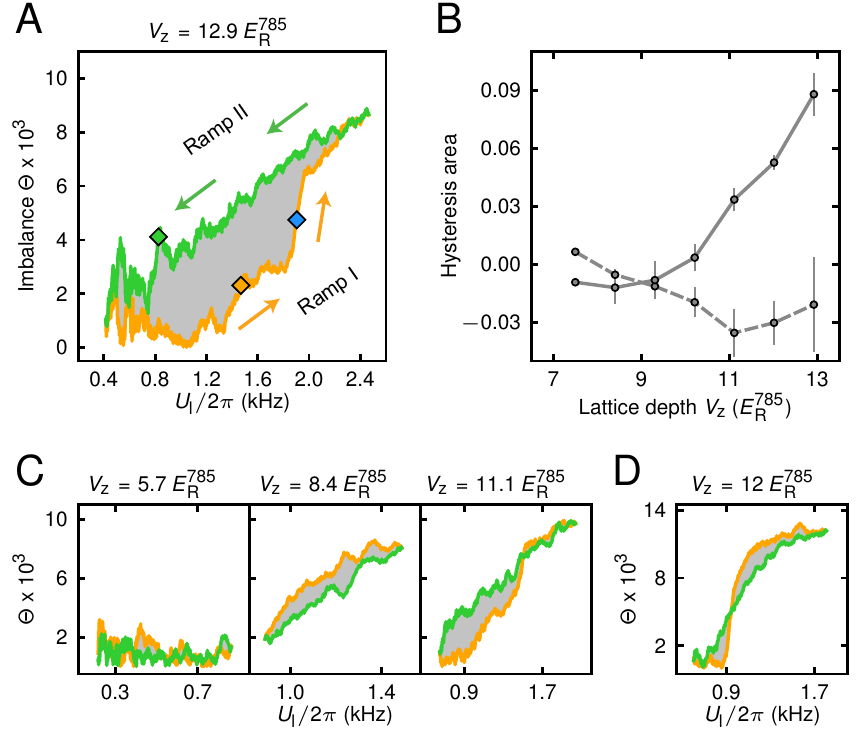}
\caption{\textit{Hysteresis Measurement.} (A) We prepare an MI, and then sweep the detuning towards cavity resonance and subsequently back to the starting point. The imbalance created during ramp~I is shown in orange and the imbalance during ramp~II is shown in green. Arrows indicate the ramp directions. We quantify the amount of hysteresis created by the area highlighted in gray. Diamonds signal where we deduce the threshold for the creation (orange) and the disappearance (green) of an imbalance \imb{}, and where the center of an imbalance jump is located (blue), see SI Appendix. (B) We study the hysteresis area as a function of the final lattice depth \Vz{}, the data is shown by the solid line. The dashed line represents the case where the $y$ lattice is switched off such as to reduce \Us{}. (C) Exemplary traces of the imbalance \imb{} as a function of \Ul{} for different lattice depths \Vz{}. These experiments were performed with $17(2)\times 10^3$~atoms at maximum lattice depths of $(\Vx{}, \Vy{}, \Vz{})=(14.5~\Er{785}, 26.2~\Er{671}, 12.9~\Er{785})$. (D) Exemplary trace with the $y$ lattice switched off. Here we prepare $15(1)\times 10^3$~atoms at $\Vx=12.4$~\Er{785} and $\Vz{}=12.0$~\Er{785}, see SI Appendix. Error bars are SD (SI Appendix).}
\label{fig:hysteresis}
\end{figure}
\section*{Metastability and Hysteresis}
A common method to probe a system for the presence of metastable states is to prepare it in a well defined state, to provide excess energy, and to observe which states it relaxes to. We accordingly implement such a \textit{Metastability Measurement} where we prepare the cloud in an MI state by slowly ramping up the lattices at an initial detuning $\Dctp{} =-50$~MHz that corresponds to a negligible strength of global-range interactions \Ul{}. Subsequently, to provide energy to the system, we quench the initial detuning within $20$~ms to a variable endpoint of \Dc{f} closer to cavity resonance. The quench increases \Ul{} while \Us{} stays unchanged. Following the quench the system evolves while all experimental parameters are kept constant. A schematic of this sequence is shown in Fig.~\ref{fig:system}C. We observe that the imbalance \imb{} rises during or after the quench until it settles at a steady-state level \imbbar{}, defined as an average over 10~ms taken 30~ms after finishing the quench (Fig.~\ref{fig:metastable}D).

Repeating the experiment, we measure the imbalance \imbbar{} as a function of the final detuning \Dc{f} (Fig.~\ref{fig:metastable}A-C). Far from resonance ($\Dctp{f}<-24$~MHz), where the strength of global-range interactions is weak, the system consistently ends up at low imbalances (orange) in an interval of $0<\imbbar{}<7 \times 10^3$~atoms. Quenching the detuning closer to resonance ($\Dctp{f}\geq -19.5$~MHz), where the strength of global-range interactions is higher, the system is never found to end up within this imbalance interval. We now observe consistently higher imbalances (green) of $\imbbar{}>7 \times 10^3$~atoms. The two well separated imbalance intervals (Fig.~\ref{fig:metastable}A) coexist for final detunings in an intermediate region ($-24 $~MHz~$\leq \Dctp{f} \leq -19.5$~MHz) where the system ends up either in a state of low or of large average imbalance \imbbar{} (SI Appendix).

We attribute the observation of two distinct imbalance distributions in our system to the existence of two metastable states. Their separation signals the presence of an energy barrier between the states which does not allow for a continuous connection between them. Our observation of a constant imbalance level after equilibration (Fig.~\ref{fig:metastable}D) shows that the final state is long-lived and hence can be either metastable or stable. Monte-Carlo simulations for the closed version of the system indeed predict metastable states \cite{Flottat2017}. We observe that this metastability is preserved in our system despite its open character due to the dissipative cavity, which could lead to a fast decay of metastable states.\\			

Metastable behavior in a many-body system is usually associated with hysteresis at phase transitions. When a control parameter is slowly varied back and forth across a critical point, the final state of the system depends on its history. The direct observation of hysteresis provides an indication for the stability of metastable states with respect to parameter changes. We perform such a \textit{Hysteresis Measurement} by preparing our system in the MI phase at a lattice depth of $\Vz{}=12.9$~\Er{785} and again at a detuning where global-range interactions are negligible. Afterwards, the detuning is swept during $80$~ms across the phase transition towards resonance (Fig.~\ref{fig:hysteresis}, ramp~I) and subsequently back to the starting point, again within $80$~ms (Fig.~\ref{fig:hysteresis}, ramp~II). We choose a detuning ramp which varies \Ul{} linearly in time, starting from $\Dctp{}=-53$~MHz to $\Dctp{}=-13$~MHz and back to $\Dctp{}=-53$~MHz, while \Us{} is kept constant (SI Appendix). During ramp~I an imbalance is created that increases with increasing \Ul{} (orange line in Fig.~\ref{fig:hysteresis}A). During ramp~II the imbalance decreases again until it fully vanishes (green line in Fig.~\ref{fig:hysteresis}A). The observed evolution of the imbalance is path dependent and describes a hysteresis loop across the phase transition. 

A natural question in our system is the connection between the strength of short-range interactions and the emergence of a hysteresis loop. We therefore repeat the experiment at different \Vz{} to vary $\Us{}/t$, where $t$ is tunneling. Sample traces are shown in Fig.~\ref{fig:hysteresis}C. We quantify the amount of hysteresis by integrating the area of imbalance with respect to \Ul{} (gray area in Fig.~\ref{fig:hysteresis}A and C). The hysteresis area is growing with increasing \Vz{}, see solid line in Fig.~\ref{fig:hysteresis}B, indicating that the metastable states become increasingly robust against a change in \Ul{}. In the case where we repeat the experiment with the $y$ lattice switched off, such as to significantly reduce $\Us{}$, we however observe barely any hysteresis area (dashed line in Fig.~\ref{fig:hysteresis}B), an exemplary trace is shown in Fig.~\ref{fig:hysteresis}D. Our findings suggest that the emergence of a hysteresis loop is linked to the system being in a regime where both interactions are strong. 

So far we neglected the influence of non-adiabaticity when crossing the phase transition point as well as heating effects. Non-adiabaticity which stems from short ramp times leads to a delayed reaction of the system with respect to a change of the detuning \cite{Klinder2015}. Consequently, during ramp~I, the imbalance build up is delayed while during ramp~II the imbalance vanishes at a later point, leading to an increase in the observed hysteresis area. Heating on the other hand leads to a reduction of the imbalance over time, resulting in a decreased hysteresis area which can thus also become negative, see Fig.~\ref{fig:hysteresis}B. The full comparison between the hysteresis area and thermodynamic states is challenging due to these effects. A detailed study of the dependence of the observed hysteresis on the ramp time is provided in the SI Appendix. Independent of the ramp time, we always observe a larger hysteresis area when \Us{} is high as compared to the case where \Us{} is reduced by switching off the $y$ lattice. Reverting the order of ramp~I and II such as to start in a CDW state would have the effect that both heating and non-adiabaticity increase the observed hysteresis area.

\begin{figure}[t]
\centering
\includegraphics[width=\columnwidth]{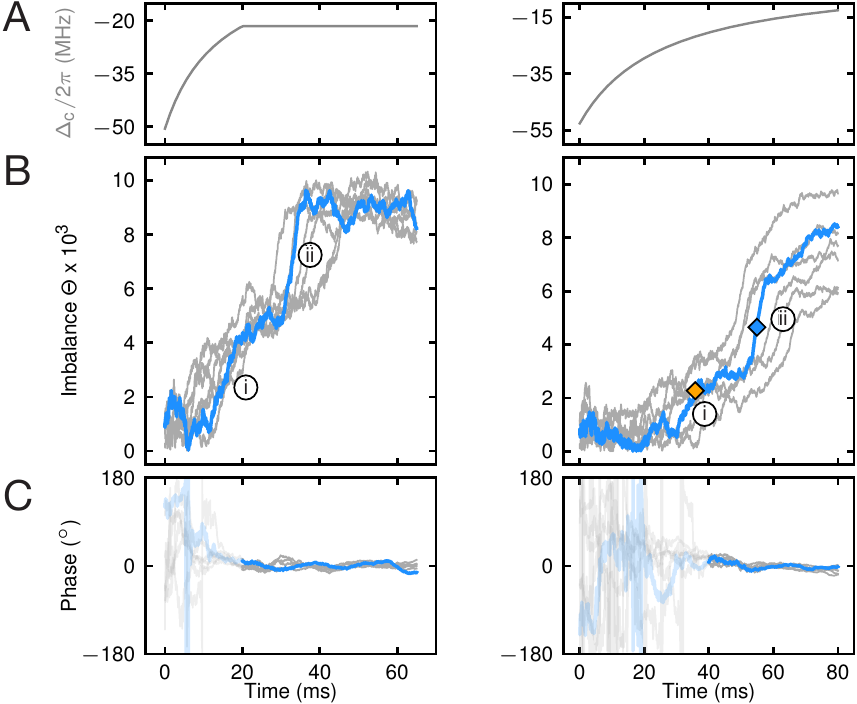}
\caption{Time traces of the dynamics of the system. Data from the \textit{Metastability Measurement} is shown in the left column, data from the \textit{Hysteresis Measurement} is shown in the right column. (A) Ramps in the detuning \Dc{}. (B) Imbalance dynamics. Starting from a state with almost zero imbalance \imb{}, we first observe a slow increase in \imb{} (i) followed by a sudden jump (ii). Left column: \textit{Metastability Measurement} data. After quenching the detuning \Dc{} in the MI phase towards cavity resonance, we hold all experimental parameters constant. We observe dynamics in the imbalance \imb{} during and after the detuning quench. The exemplary trace of \imb{} as a function of time at a final detuning of $\Dctp{f}=-21$~MHz is shown in blue, while several repetitions of the experiment at $\Dctp{f}=-23\textsf{ to }-20$~MHz are shown in grey. Right column: \textit{Hysteresis Measurement} data. We sweep the detuning \Dc{} within $80$~ms from the MI phase towards cavity resonance. An exemplary trace of \imb{} as a function of time is shown in blue where we observe dynamics in the imbalance during the sweep. Multiple repetitions of the experiment with the same parameters are shown in grey, here $\Vz=12.9$~\Er{785}. Diamonds signal where we deduce the threshold for the creation of an imbalance \imb{} (orange), and where the center of the imbalance jump (ii) is located (blue), see SI Appendix. (C) Phase of the light field indicating a broken $\mathbb{Z}_2$-symmetry. We observe a constant phase after an imbalance is created throughout the slow increase (i) and jump (ii) in \imb{}. In the shaded region, the signal is dominated by technical noise due to low photon flux.}
\label{fig:imbalance_dynamics}
\end{figure}
\section*{Imbalance dynamics}
Our findings in the previous two experiments, the \textit{Metastability Measurement} and the \textit{Hysteresis Measurement}, are based on changes of the imbalance \imb{} when varying the detuning \Dc{} in time. Such a change in the imbalance corresponds to a reordering of the atomic density distribution via tunneling in the lattice potential. Our real-time access unveils non-trivial dynamics of the imbalance in the same data. We observe an initial imbalance build up (i) followed by a fast jump (ii), see Fig.~\ref{fig:imbalance_dynamics}. Both features are present in the case of a detuning quench and the case of a slow detuning ramp. To be independent of the quench time, we post-select the quench data based on the condition that the imbalance jump (ii) happens after experimental parameters are kept constant (SI Appendix). From this data we measure a height of the jump of $\Delta\imb{}=3.5(9)\times 10^3$~atoms and an upper bound of the duration $\Delta T$ of $4.3(6)$~ms (SI Appendix). It is comparable to the tunneling time in a double well along the $x$[$z$] direction of $11.8[3.1]$~ms, defined as $\pi/(2 \sqrt2 \, t_{x[z]})$ (SI Appendix). We interpret this jump as a collective tunneling of several thousand atoms, a possible microscopic description of this process is given in the following section. The timescale of the initial imbalance build up (i) depends on the ramp time, while the jump (ii) has a comparable duration in all datasets. 

\begin{figure}[!th]
\centering
\includegraphics[width=\columnwidth]{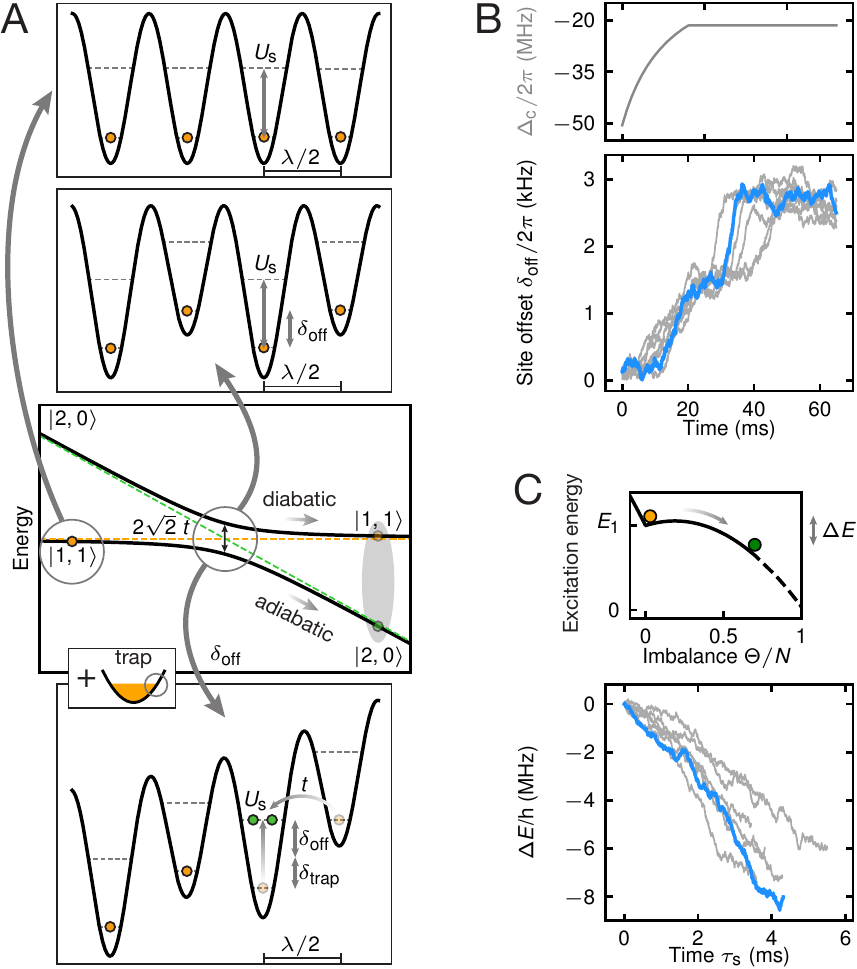}
\caption{Microscopic dynamics and energy redistribution of the system. (A) Microscopic description of the system dynamics following the detuning quench, in terms of a Landau-Zener transition. One-dimensional lattice potentials are shown for a normal lattice (top), a dynamic superlattice with site offset \del{off} generated by superfluid surface atoms (middle), and a tilted dynamic superlattice with spatially varying site offset $\del{off}+\del{trap}$ as encountered at the edge of the harmonic trap (bottom). Colored circles represent atoms in the states $\left|1,1\right>$ (orange) or $\left|2,0\right>$ (green). Resonant nearest neighbor tunneling is allowed when the site offset $\del{off}+\del{trap}$ equals the short-range interaction strength \Us{}. (B) Dynamics of the site offset \del{off} in the \textit{Metastability Measurement}. (C) Top panel: Sketch of the excitation energy of the bulk atoms. Superfluid surface atoms add a symmetry breaking field to the toy model. During the imabalance jump (ii), the highly excited system reduces the initial excitation energy $E_1$ via an avalanche of inherently non-adiabatic Landau-Zener transitions by an amount of \dE{}. Colored circles represent the state of the system, where the MI state (orange) results from all bulk atoms in the $\left|1,1\right>$ state, and the CDW state (green) from atoms being in a superposition of $\left|1,1\right>$ and $\left|2,0\right>$ states. Accordingly, the relative imbalance saturates at $\imb{}/N<1$, indicated by the dashed line. Bottom panel: Reduction of \dE{} as a function of time $\tau_{\text{s}}$ during the imbalance jump (ii). (B-C) Exemplary traces use the same data as shown in Fig.~\ref{fig:imbalance_dynamics}. \del{off} and \dE{} are inferred from the photon flux leaking from the cavity.}
\label{fig:energy_dynamics}
\end{figure}

In contrast to our toy model, the experimental system is at non-zero tunneling, at finite temperature, and exposed to a harmonic trapping potential. Accordingly we expect the MI, in which we initially prepare the system, to form a wedding cake structure consisting of an insulating bulk surrounded by superfluid shells at the surface. Such an inhomogeneous finite size system can exhibit a first order phase transition of the bulk material (the MI), which is triggered by a second order phase transition that took place previously on the system's surface \cite{Lipowsky1983, Lipowsky1987}. The superfluid surface atoms possess a higher mobility than the insulating bulk \cite{Hung2010}. When the detuning \Dc{} is swept towards cavity resonance, these atoms can gradually create an imbalance once global-range interaction overcome kinetic energy and the trapping potential. The emerging imbalance breaks the discrete $\mathbb{Z}_2$-symmetry of the CDW state, indicated by a well defined and constant phase of the measured light field \cite{Baumann2011a}, shown in Fig.~\ref{fig:imbalance_dynamics}. We attribute the initial imbalance increase (i) to a rearrangement of surface atoms. From the experimental parameters of the \textit{Metastability Measurement}, we theoretically estimate a number of surface atoms of $N_{\text{surf}}\approx {(4-8)}\times 10^3$ (SI appendix), which is in agreement with the initial imbalance increase (i). Photons scattered at these atoms into the cavity mode generate an energy offset \del{off} between even and odd sites, see Fig.~\ref{fig:energy_dynamics}B. This offset eventually drives the bulk system from a metastable MI to a CDW state, which we link to the fast imbalance jump (ii). However, we do not observe an imbalance jump when ramping the detuning back to the starting value in ramp~II (Fig.~\ref{fig:hysteresis}), which we mainly attribute to the cloud being heated.

\section*{Microscopic dynamics and energy redistribution during the imbalance jump}
A simplified microscopic picture of the imbalance dynamics following the detuning quench is sketched in Fig.~\ref{fig:energy_dynamics}A, where the system is broken into a collection of coupled double wells. In the initial MI state, bulk atoms occupy both sites of each double well. This state is labeled $\left|1,1\right>$, where $\left|n_{\text{e}},n_{\text{o}}\right>$ denotes the filling on the even and odd sites, respectively. Here, on-site interactions of strength $\Us{}/2\pi=2.2(1)$~kHz provide an energy barrier for neighboring atoms, thus suppressing tunneling into a $\left|2,0\right>$ state. The barrier softens but persists as surface atoms generate an imbalance \imb{} and a site offset \del{off}. Monitoring the flux of photons leaking from the cavity, we observe $\del{off}/2\pi=1.6(2)$~kHz just before the imbalance jump (ii) happens (SI Appendix). The harmonic trapping potential causes an additional site offset of $0 \text{~kHz} \leq \del{trap}/2\pi \leq \delta_{\text{trap}}^{\text{max}}/2\pi = 0.6 \text{~kHz}$, increasing from the center outwards. When $\del{off}+\delta_{\text{trap}}^{\text{max}} \approx \Us{}$, the outermost bulk atoms start resonantly tunneling to their neighboring lattice sites. They further increases \imb{} and \del{off}, successively allowing more and more atoms to resonantly tunnel. The imbalance jump (ii) thus results from an avalanche of resonant tunneling processes of bulk atoms which only stops once $\del{off}-\delta_{\text{trap}}^{\text{max}} > \Us{}$. Indeed, we find $\del{off}/2\pi=2.7(3)$~kHz at the end of the jump.

We describe each resonant tunneling process by a Landau-Zener transition, shown in Fig.~\ref{fig:energy_dynamics}A. The $\left|1,1\right>$ and $\left|2,0\right>$ states are coupled with strength $\sqrt{2}t$, where the tunneling $t$ is bosonically enhanced by a factor of $\sqrt{2}$. We find an upper bound for the probability of adiabatic Landau-Zener transfer of about $60$~\%, which is determined by the measured rate of change of \del{off} during the imbalance jump, shown in Fig.~\ref{fig:energy_dynamics}B. As all experimental parameters are held constant after the quench, the site offset \del{off} is solely tuned by the reordering atoms. The timescale and (non-)adiabaticity of the Landau-Zener transitions is thus inherently determined by the system evolving non-linearly due to the presence of the global-range interactions.\\

At the beginning of the imbalance jump (ii), the ground state of the system is the CDW state. The bulk is however still in the MI state, which is now a highly excited state of energy $E_1$. During the imbalance jump (ii) each double well in the bulk evolves via non-adiabatic Landau-Zener transfers to a superposition of $\left|1,1\right>$ and $\left|2,0\right>$ states. On top of the imbalance created previously by superfluid surface atoms, the redistributing bulk increases the imbalance further, allowing the system to lower the excitation energy by $\dE{}$. We infer $\dE{}=7.7(2.1)$~MHz from the imbalance jump (ii) in the \textit{Metastability Measurement} (SI Appendix), see Fig.~\ref{fig:energy_dynamics}C. This process is sketched using our toy model, where a symmetry breaking field is present due to the imbalance created by superfluid surface atoms.

In order to study the energy budget of the system we consider two scenarios. If the system was closed, the total energy could not change, and the reduction in excitation energy \dE{} would be balanced by an increase in kinetic energy of the system. Since our system is inherently open, the energy could also be dissipated by leaking cavity photons. We make use of the spectrum of these photons to distinguish the two cases. We estimate the number of scattered photons during the imbalance jump (ii) to be about $12(3)\times 10^{3}$ (SI Appendix), where each photon would have to dissipate at least $0.6(3)$~kHz of energy. This would leave a notable signature in the photon spectrum, which is not observed. While our heterodyne detection cannot rule out processes where only few photons dissipate all the energy, such a collective scattering process seems unlikely. Hence we conclude that the excitation energy released during the jump (ii) is transformed into kinetic energy of the system.

\begin{figure}[t]
\centering
\includegraphics[width=\columnwidth]{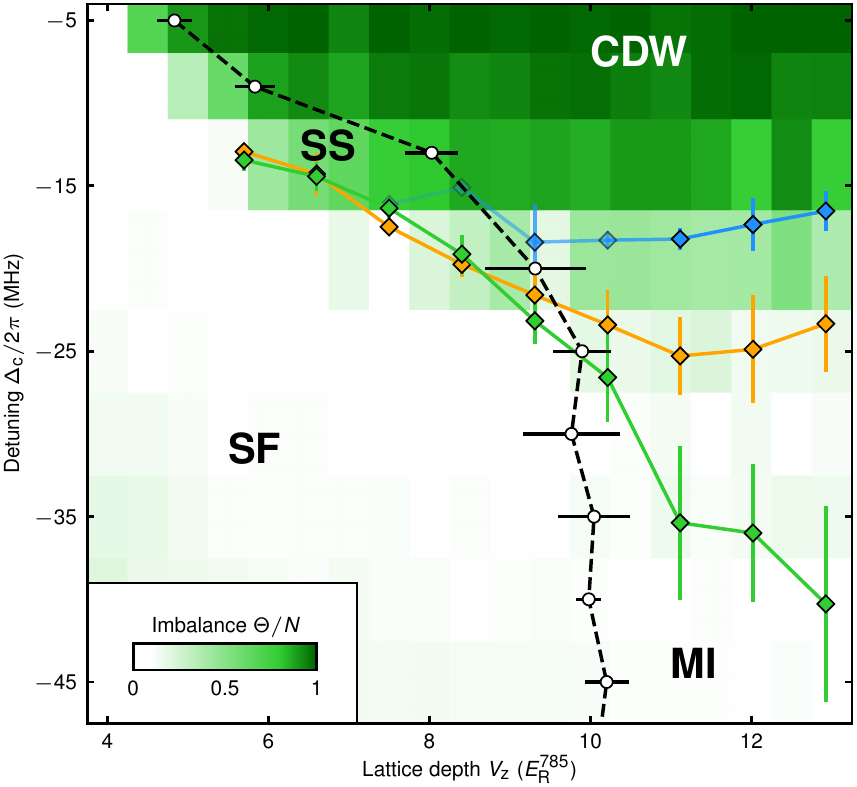}
\caption{Previously extracted transition points superimposed on a phase diagram of the system. Results from the \textit{Hysteresis measurement:} Orange and green diamonds indicate the thresholds where an imbalance is created and where it vanishes during detuning ramps, respectively. The center of the imbalance jump is shown in blue, where transparency indicates the probability of occurrence of the jump. For details on the measurement of the phase diagram, see SI Appendix. White data points and the associated black dashed line indicated the loss of coherence, from left to right, which we infer from the measured BEC fraction, and green tiles indicate states with non-zero imbalance. We identify a superfluid (SF), a lattice supersolid (SS), a Mott-insulator (MI) and a charge-density wave (CDW) phase. This experiment was performed with $16(1)\times 10^3$~atoms at maximum lattice depths of $(\Vx{}, \Vy{}, \Vz{})=(15.7~\Er{785}, 26.2~\Er{671}, 12.9~\Er{785})$. For further details see \cite{Landig2016} and SI Appendix. Error bars are SD (SI Appendix).}
\label{fig:phase_diagram}
\end{figure}
\section*{Phase diagram}
The observation of metastable states, a coexistence of phases and a jump in the order parameter are typical features of first order phase transitions. We thus want to relate our observations to a phase diagram of the system measured as in \cite{Landig2016}, see Fig.~\ref{fig:phase_diagram}. Here, we superimpose the thresholds extracted in the \textit{Hysteresis Measurement} on the phase diagram. 

The threshold for the creation of an imbalance (orange diamonds) coincides with the appearance of an imbalance in the phase diagram (green tiles). The center position of the fast jump (blue diamonds) is located within a region of intermediate imbalance present in the phase diagram at $\Dctp{}\approx -20$~MHz (light green tiles). The threshold for the disappearance of an imbalance (green diamonds) extends deep into the MI region (white tiles). The associated blue and green lines enclose an area where the MI and the CDW phases can coexist and where hysteresis is observed. In addition, we find the parameter regime where the system can fall into either of the two final states in the \textit{Metastability Measurement} (Fig.~\ref{fig:metastable}A) to lie close to the blue line (Fig.~\ref{fig:phase_diagram}). 

\section*{Conclusion and Outlook}
Using the unique real-time access of our experiment, we observed long-lived metastable phases and hysteretic behavior at a first-order quantum phase transition between an MI and a CDW phase. Owing to the non-linearity stemming from the global-range interactions, the system develops its own timescale when quenched across the phase transition. The resulting dynamics of spatially reordering atoms points to an avalanche of resonant tunneling processes taking place, which render the transition out of the metastable state inherently non-adiabatic. The observed lack of energy dissipation during the transition poses questions on the thermalization of the final state. Our work provides a novel approach to study dynamics and thermalization processes in open quantum many-body systems.\\

\begin{acknowledgments}
We acknowledge insightful discussions with Frederik G\"org, Katrin Kr\"oger, Gabriel T. Landi, Giovanna Morigi, Helmut Ritsch, Andr\'{e} Timpanaro, P\"aivi T\"orm\"a, Sascha Wald, and Wilhelm Zwerger. We acknowledge funding from Synthetic Quantum Many-Body Systems (a European Research Council advanced grant) and the EU Collaborative Project TherMiQ (Grant Agreement 618074), and also SBFI support for Horizon2020 project QUIC, and SNF support for NCCR QSIT and DACH project ‘Quantum Crystals of Matter and Light’.
\end{acknowledgments}

\clearpage
\newpage 

\onecolumngrid
\appendix
\makeatletter
\setcounter{section}{0}
\setcounter{subsection}{0}
\setcounter{figure}{0}
\setcounter{equation}{0}
\renewcommand{\theequation}{S\arabic{equation}}
\renewcommand{\thefigure}{S\arabic{figure}}
\renewcommand{\bibnumfmt}[1]{[S#1]}
\renewcommand{\citenumfont}[1]{S#1}

\section*{SI Appendix}
\subsection{Lattice calibrations}
We calibrate the lattice depth along the $x$ direction by amplitude modulation spectroscopy observing the position of the lowest three Bloch bands \cite{Stoferle2004_sup}. The lattice depths along the $y$ and $z$ direction are calibrated via Raman-Nath diffraction \cite{Morsch2006_sup}. The lattice depths are calibrated separately for each experiment, and we obtain the following parameters. In the \textit{Metastability Measurement}, $\Vx{}=1.56(5)\times \Vz{}$, $\Vy{}=30.7(1.6)$~\Er{671}, and $\Vz{}=11.1(7)$~\Er{785}. In the \textit{Hysteresis Measurement}, $\Vx{}=1.12(3)\times \Vz{}$, $\Vy{}=26.2(1.1)$~\Er{671} when the $y$ lattice is present and $\Vx{}=1.02(4)\times \Vz{}$ when the $y$ lattice is switched off, and $\Vz{}$ ranging from $5.7(4)$~\Er{785} to $12.9(2)$~\Er{785}. In the phase diagram measurement, $\Vx{}=1.27(11)\times \Vz{}$, $\Vy{}=26.2(1.1)$~\Er{671}, and $\Vz{}$ ranging from $4.0(5)$~\Er{785} to $12.9(2)$~\Er{785}. Errors on the lattice depths in the $y$ and $z$ direction incorporate uncertainties from the calibration and residual offsets on the photodiodes.

\subsection{Detuning calibrations}
\begin{figure}[h]
\centering
\includegraphics[width=87mm]{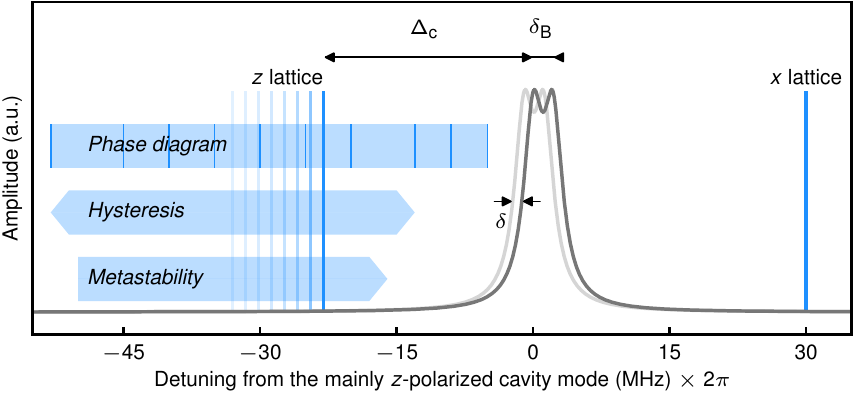}
\caption{Various detunings used in the experiment. The resonances of the two linearly polarized TEM$_{00}$ modes of the empty cavity are shown by the dark grey line, their resonance frequencies are separated by $\ddtp{B}=2.2$~MHz due to birefringence. The full width at half maximum (FWHM) of each resonance is $2\kappa/2\pi=2.5$~MHz. Coupling of atoms to the cavity shifts the cavity resonance by the dispersive shift $\delta$ down in frequency (light grey line). The $z$ lattice is detuned by a variable amount of \Dc{} from the lower lying resonance frequency of the mainly $z$-polarized mode of the empty cavity, shown by the vertical blue line on the left half of the figure, where the neighboring faint blue lines illustrate the scan direction. The $x$ lattice is detuned by $2\pi\times 30$~MHz from the same mode, shown by the vertical blue line on the right. Horizontal arrows depict the scan directions and ranges of the different experiments, and the small vertical blue lines indicate where the phase diagram data is taken.} 
\label{fig:detunings}
\end{figure}

The BEC couples to two linearly polarized TEM$_{00}$ eigenmodes of the cavity, which are tilted by $\alpha=22^{\circ}$ with respect to the $y$ and $z$ axis. The resonance frequencies of the eigenmodes are separated due to birefringence by $\ddtp{B}=2.2$~MHz. The detuning \Dc{} refers to the lower lying resonance frequency of the mainly $z$-polarized mode, and the $x$ lattice is detuned by $2\pi\times 30$~MHz from this mode, see Fig.~\ref{fig:detunings}. In every experimental repetition, after atomic absorption pictures are taken, we scan the frequency of the $x$ lattice across the cavity resonance and fit the resulting photon signal with a Lorentzian. We deduce a standard deviation of \Dctp{} of $0.3$~MHz.

\section{Magnetic fields and gradients}
We apply a magnetic gradient field levitating the atomic cloud. In addition, we operate the experiment at a magnetic offset field large enough to achieve a good separation between the atomic hyperfine levels such as to avoid Raman-assisted spin-flips induced by the presence of the lattices and the cavity. We use a magnetic field of $B\approx 130$~G oriented along the $z$ axis and obtain a Zeeman-splitting of about $\Delta E/h \approx 90$~MHz, well above the maximum cavity detuning of $\Dctp{max}=-53$~MHz. 

\section{Trapping frequencies}
In our system, the cloud is magnetically levitated and subject to a crossed far off-resonant dipole trap. In the absence of optical lattices, we calculate the trapping frequencies in all three directions and find $(\omega_x, \omega_y, \omega_z) = 2\pi\times(96, 38, 49)$~Hz, respectively. When we include a $671$~nm blue-detuned $y$ lattice of depth $\Vy{}=30$~\Er{671}, which is comparable to our experimental parameters, together with an increased dipole trap depth, we calculate trap frequencies of $(\omega_x, \omega_y, \omega_z) = 2\pi\times(116, 38, 67)$~Hz. In the case of lattice depths comparable to the maximum lattice depths used in the experiment ($\Vx{}=\Vz{}=14$~\Er{785}, $\Vy{}=30$~\Er{671}) we calculate trap frequencies of $(\omega_x, \omega_y, \omega_z) = 2\pi\times(219, 221, 193)$~Hz. Deconfinement due to changing zero-point energies is taken into account. We compare our calculations with experimental data and find good agreement. We estimate an error of about $10~\%$ resulting primarily from uncertainties in the determination of beam waists at the position of the atoms.

\section{Extraction of the even-odd particle imbalance \imb{} and site offset \del{off} from the measured photon flux}
We obtain the imbalance \imb{} from the mean intra-cavity photon $n_{\textsf{ph}}$ number via
\begin{equation}
\imb{} = \Big| \sum_{i \in e}{\left\langle\hat{n}_i \right\rangle} - \sum_{i\in o}{\left\langle\hat{n}_i\right\rangle}\Big| = \sqrt{n_{\mathrm{ph}}\, \frac{\Dc{2}}{\eta^2 M_0^2}}\frac{1}{F(\Dc{})},
\label{eq:Theta}
\end{equation}
with 
\begin{align}
F(\Dc{}) &= \Dc{}   \Big | \frac{\cos^2\alpha}{\Dc{'}-\dd{B}+i\kappa}   +  \frac{\sin^2\alpha}{\Dc{'}+i\kappa}    \Big|  
\stackrel{|\Dc{}| \gg \kappa, |\delta |,\dd{B}}{\approx} 1
\end{align}
$F(\Dc{})$ takes into account the two linearly polarized $\mathrm{TEM}_{00}$ eigenmodes of the cavity. The effective two-photon Rabi frequency is given by $\eta/2\pi= 2.99\sqrt{\Vz{}/\hbar}\sqrt{\mathrm{Hz}}$, the spatial overlap of the interference lattice provided by the cavity mode and the $z$ lattice with the Wannier-function $W_i(x, z)$ of an atom localized at lattice site $i$ is given by $M_0 = \int\int \mathrm{d}x \, \mathrm{d}z \, W_i^{*}(x,z) \cos{\left(kx\right)} \cos{\left(kz\right)} W_i(x,z)$, the cavity decay rate is $\kappa/2\pi=1.25$~MHz, and $\Dc{'}=\Dc{}-\delta$ takes into account the dispersively shifted cavity resonance, where $\delta$ corresponds to the dispersive shift with a maximum shift per atom of $U_0/2\pi=-56.3$~Hz for each of the two cavity modes. A moving average of window size $4$~ms is used on all photon data except for the phase diagram in Fig.~\ref{fig:phase_diagram} where the window size is $10$~ms. Note: Technical noise on the photon detector is converted into an imbalance \imb{}. Due to the dependence of \imb{} on \Dc{} and \Vz{} the background noise causes a noticable signal far from cavity resonance, and contributes to the small but non-zero imbalance visible on the left side of Fig.~\ref{fig:metastable}D, Fig.~\ref{fig:hysteresis}A and C-D, Fig.~\ref{fig:imbalance_dynamics}B, and to the imbalance visible in the lower left corner of Fig.~\ref{fig:phase_diagram}. For further details see \cite{Landig2016_sup}.

The energy offset \del{off} between even and odd sites is related to the strength of the dynamic checkerboard lattice depth formed by the $z$ lattice and the light scattered into the cavity. It is defined as
\begin{equation}
\del{off} = 4 \eta M_0\sqrt{n_{\mathrm{ph}}}
\end{equation} 

\section{Strength of effective atom-atom interactions of global-range}
Taking both cavity modes into account, \Ul{} is given by
\begin{align}
&\Ul{} = -K |\eta M_0|^2 \left[  \frac{(\Dc{'}-\dd{B})\cos^2\alpha}{\left(\Dc{'}-\dd{B}\right)^2 + \kappa^2}  +   \frac{\Dc{'}\sin^2\alpha}{\Dc{'2} + \kappa^2}  \right]\nonumber \\
&\stackrel{\textsf{$|\Dc{}| \gg \kappa, |\dd{} |, \dd{B}$}}{\approx}  -K |M_{0}|^{2} \frac{\eta^2}{\Dc{}}  \propto  \frac{\Vz{}}{\Dc{}}s.
\label{eq:Ul}
\end{align}
We take the number of lattice sites to be the number of atoms, $K=N$, for details see \cite{Landig2016_sup}.

\section{Derivation of the extended Bose-Hubbard toy model}
Our system is well described by a Bose-Hubbard Hamiltonian with additional global-range interactions of the form \cite{Landig2016_sup, Dogra2016_sup}:
\begin{align}
\frac{\Ham{}}{\hbar}  =  -t& \sum\limits_{<i, j>} \big(\hat{b}_{i}^{\dagger}\hat{b}_{j}+h.c. \big)+
\frac{\Us{}}{2} \sum_{i\in \textsf{e},\textsf{o}} \hat{n}_{i}\left(\hat{n}_{i}-1\right)
-\sum_{i\in \textsf{e},\textsf{o}} V_{i} \hat{n}_{i}
- \frac{\Ul{}}{K}\left(\sum\limits_{i\in \textsf{e}}{\hat{n}_{i}} - \sum\limits_{i\in \textsf{o}}{\hat{n}_{i}}\right)^{2}
\label{eq:extBH_ham}
\end{align}
where $t$ is the nearest neighbor tunneling rate, $V_{i}$ is the site dependent harmonic trapping potential, $\hat{b}_i$ and $\hat{b}^{\dagger}_i$ are the bosonic annihilation and creator operators at site $i$, and $\hat{n}_i = \hat{b}^{\dagger}_i\hat{b}_i$ is the corresponding number operator. In our toy model, we assume the limit of zero tunneling and neglect the harmonic trapping potential. We consider a total of $N$ atoms to be distributed among a fixed number of lattice sites $K=N$, independent of the strength of global-range interactions \Ul{}. This is experimentally realistic for deep lattices where an atomic wavepacket cannot spread more than a few lattice sites during the experiment due to very small tunneling $t$.

In the limit of $t=0$, the eigenstates of the system are the number states, and we can replace all the number operators in \eqref{eq:extBH_ham} by the corresponding average values, $\langle\hat{n}_{i}\rangle = {n}_{i}$. We introduce the imbalance \imb{} as 
\begin{equation}
\imb{} = \langle \hat{\Theta}  \rangle = \Big\langle \sum\limits_{i\in \textsf{e}}{\hat{n}_{i}} - \sum\limits_{i\in \textsf{o}}{\hat{n}_{i}} \Big\rangle = \sum\limits_{i\in \textsf{e}}{{n}_{i}} - \sum\limits_{i\in \textsf{o}}{{n}_{i}}
\end{equation}
The atomic configuration of least energy for a given imbalance \imb{} corresponds to part of the system being in a CDW state, namely the fraction $f_{\textsf{CDW}}=|\imb{}|/K$, while all other atoms are in an MI state. The CDW is characterized by $n^{\textsf{even}}_{i}=2$ and $n^{\textsf{odd}}_{i}=0$ for even and odd sites, respectively, while $n_i=1$ on all lattice sites in the MI state. The energy of such a state is:
\begin{equation}
\langle \Ham{} \rangle = N\varepsilon = \frac{1}{2}f_{\textsf{CDW}}\Us{}K - \frac{\Ul{}}{K} \imb{}^2 = \frac{1}{2}\frac{|\imb{}|}{K}\Us{}K - \frac{\Ul{}}{K} \imb{}^2
\end{equation}
With $N=K$, we obtain
\begin{equation}
\varepsilon = \frac{1}{2}\Us{}\frac{|\Theta|}{N} - \Ul{} \Big(\frac{|\Theta|}{N}\Big)^2
\end{equation}

The system changes its ground state from an MI state with no imbalance to a CDW state with maximum imbalance ($|\imb{}|/N = \pm 1$) at $\UlUs{} = 1/2$, see Fig.~\ref{fig:system}A. We use this critical point to calculate the energy barrier per particle between the MI and the CDW state which is defined as $E^{\textsf{barrier}}=E_m-E_g$. Here $E_g=0$ is the ground state energy and $E_m=\Us{}/8$ is the maximum energy as a function of $\imb{}$ at $|\imb{}|/N = 1/2$. For $(\Vx{}, \Vy{}, \Vz{})=(13~\Er{785}, 26~\Er{671}, 13~\Er{785})$. We obtain an energy barrier of $E^{\textsf{barrier}}/h=260$~Hz which is much larger than the single particle tunneling rate $t/2\pi=46$~Hz.

The presence of a trapping potential can lower the energy of a state of intermediate imbalance and possibly reduce the height of the energy barrier between the MI and CDW phases. This reduction in energy can be as large as $600$~Hz at the edge of the central 2D layer. Assuming such a situation to be present everywhere in the system, the energy barrier per particle is reduced to about $E^{\textsf{barrier}}/h=180$~Hz, which is still significantly larger than the tunneling rate $\sqrt{2}t$ \cite{Panas2017_sup}. 

\section{Calculation of atomic density distributions}
\textit{Number of 2D layers:} We calculate the number of 2D layers based on the measured atom number and the calculated trap frequencies, following \cite{Pedri2001_sup}. Since the lattice along the $y$ direction is very deep we assume the atom number in each 2D layer to be fixed.

\textit{Maximum lattice filling:} Following \cite{Dhar2011_sup}, we calculate the atomic density as a function of $\mu / \Us{}$ and $t/ \Us{}$ in the grand canonical ensemble, where $\mu$ is the chemical potential. Using the local density approximation and calculated trapping frequencies, we obtain the full density distribution of the atomic cloud which is used to estimate the maximum filling $n_i$.

\textit{Number of surface atoms $N_{\text{surf}}$:} In the \textit{Metastability Measurement} the 2D lattice has different strengths in the $x$ and $z$ directions ($\Vx{}=17.3$~\Er{785}, $\Vz{}=11.1$~\Er{785}). We estimate $N_{\text{surf}}$ from the calculated atomic density distribution in a balanced 2D square lattice around the average lattice depth $\overline {V}=\frac{1}{2}(\Vx{}+\Vz{})$. We obtain $N_{\text{surf}}=(4-8)\times 10^3$ atoms at $\overline {V}=(15-13)$~\Er{785}, respectively.

\section{Evaluation of the \textit{Metastability Measurement}}
The data is taken in a range of final detunings of $-36 \leq {\Dctp{f}} \leq -16$~MHz with an interval of $0.5$~MHz, amounting to a total of $41$ datasets. For every $\Dc{f}$, the experiment is repeated $13-22$ times. In each repetition we start with a detuning ramp in the time interval $0<T<20$~ms, followed by a free-evolution at $20 \leq T < 70$~ms. The imbalance \imbbar{} is obtained as the mean of the imbalance \imb{} in the time interval $50<T<60$~ms. The two distinct imbalance distributions are highlighted by coloring data with $0 < \imbbar{} < 7 \times 10^3$~atoms in orange and data with $\imbbar{} > 7 \times 10^3$~atoms in green, see Fig.~\ref{fig:metastable}.

At each final detuning \Dc{f}, we take the mean and standard deviation of data in the orange and green region separately, and we obtain Fig.~\ref{fig:metastable}A. In order to quantify the gap between the two states, we consider the final detuning region where we find states with both small and large imbalance simultaneously, i.e. $-24$~MHz $ \leq {\Dctp{f}} \leq -19.5$~MHz. We consider data above and below $\imbbar{}=7 \times 10^3$~atoms separately and take the mean and standard deviation. The difference defines the gap between the two states, which has a height of $5.2(1.4) \times 10^3$~atoms. In another representation of the same data, we split the imbalance data of each \Dc{f} into $22$ bins of binsize $700$ atoms and construct a histogram as a function of \imbbar{} and \Dc{f}, see Fig.~\ref{fig:metastable}B. In order to obtain mean counts $\overline{\textsf{Counts}}$ as shown in Fig.~\ref{fig:metastable}C, we generate a histogram with respect to \imbbar{} of data in the orange region of Fig.~\ref{fig:metastable}B, where we normalize counts by the respective sample size ($492$), and we repeat this procedure for data in the green region which has a sample size of $181$. This way the obtained histogram becomes independent of the exact sample size in each state, as the sample size is sensitive to the scan region of final detunings.

\section{\textit{Hysteresis Measurement}: Lattice and detuning ramps}
\begin{figure}[ht]
\centering
\includegraphics[width=87mm]{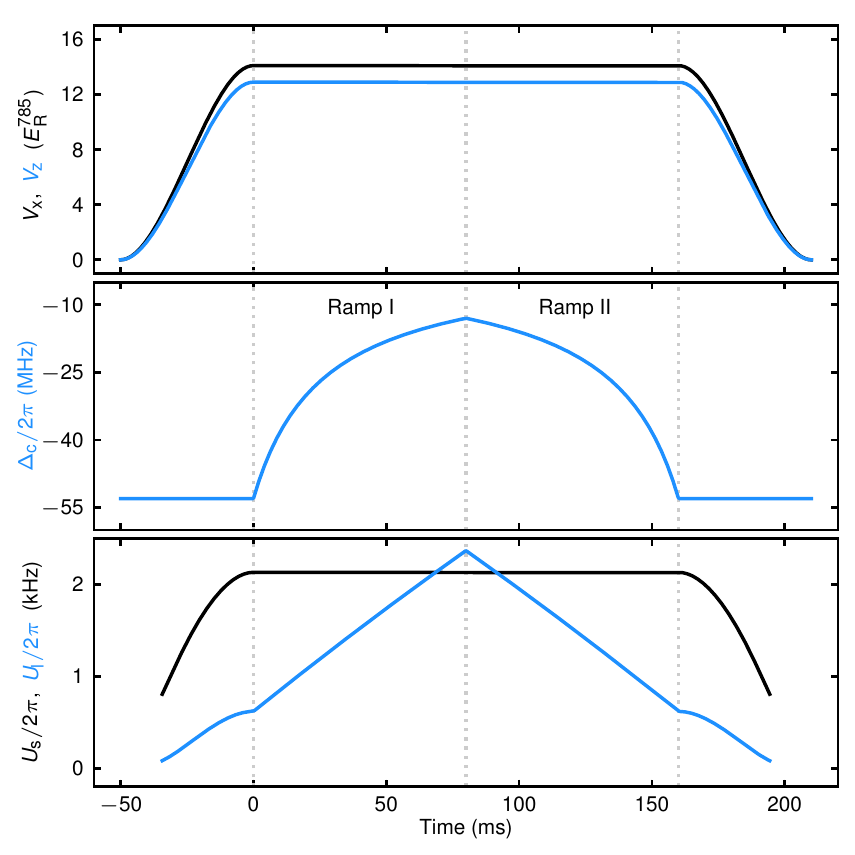}
\caption{Temporal sequence of lattice amplitude and detuning ramps in the \textit{Hysteresis Measurement}. Time $T=\left[-50, 0\right]$~ms: The square lattice in the $x-z$ direction is ramped to a depth of $(\Vx{}, \Vz{})$ (top panel)) at a constant detuning \Dc{} (middle panel), predominantly increasing \Us{} but also \Ul{} (bottom panel). $T=\left[0, 80\right]$~ms: The detunig \Dc{} is ramped towards resonance (ramp~I) such as to vary \Ul{} linearly in time while \Us{} is kept constant. $\left[80, 160\right]$~ms: The ramp in the detunig is inverted and \Dc{} is brought back to the starting point (ramp~II). $\left[160, 210\right]$~ms: The square lattice in the $x-z$ direction is ramped down again.}
\label{fig:exp_sequence}
\end{figure}
The BEC is initially prepared in a crossed far off resonant dipole trap. Then a strong $y$ lattice is ramped within $100$~ms to a final depth of \Vy{}, where the ramp follows an S-shape of form $V(T) = V_0\left[3\left(\frac{T}{T_0}\right)^2 - 2\left(\frac{T}{T_0}\right)^3 \right]$. Here $V_0$ is the final lattice depth, $T$ is time, and $T_0$ is the total duration of the ramp. The $y$ lattice cuts the cloud into weakly coupled 2D-layers. The subsequent sequence of amplitude and detuning ramps is shown in Fig.~\ref{fig:exp_sequence}. First, the square lattice in the $x-z$ direction is applied using another S-shaped amplitude ramp of $50$~ms duration, finishing at depths \Vx{} and \Vz{}. Then the $z$ lattice detuning \Dctp{} is swept from $-53$~MHz to $-13$~MHz within a variable time of $\tau=(30-150)$~ms using a ramp which varies \Ul{} linearly in time. The ramp has the form $\Dc{}(T)=\left[  \left( \frac{1}{\Dc{}(\tau)} - \frac{1}{\Dc{}(0)}\right )\frac{T}{\tau} + \frac{1}{\Dc{}(0)}  \right]^{-1}$, where $\Dc{}(0)$ and $\Dc{}(\tau)$ represent the initial and final detuning, respectively. Subsequently, the detuning \Dctp{} is swept back to $-53$~MHz, using an inverted ramp of the same duration. Finally, the square lattice is ramped down within $50$~ms using another S-shaped ramp.

\section{Hysteresis loops: Data evaluation and comparison of different ramp times}
\begin{figure}[ht]
\centering
\includegraphics[width=87mm]{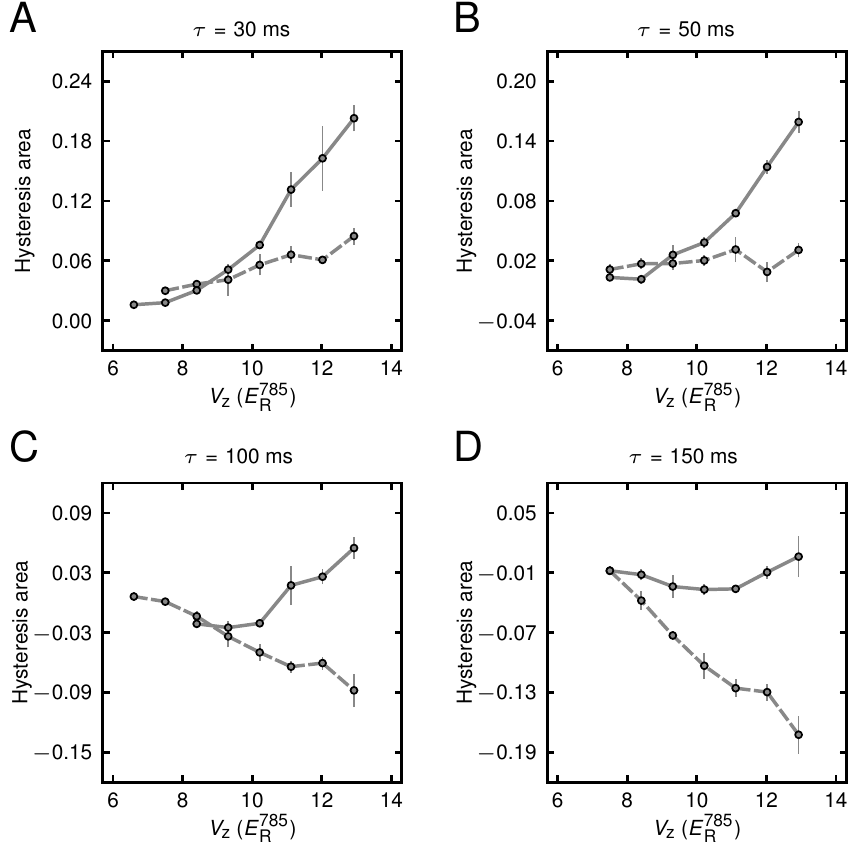}
\caption{Hysteresis area as a function of the ramp time $\tau$. We obtain the hysteresis area for ramp times (A) $\tau=30$~ms, (B) $\tau=50$~ms, (C) $\tau=100$~ms, and (D) $\tau=150$~ms in the same way as in Fig.~\ref{fig:hysteresis}B. Solid lines represent the case where lattices in all three directions are applied, while dashed lines represent the case where the $y$ lattice is switched off such as to reduce \Us{}. Data points represent statistical means and errors are SD.} 
\label{fig:timesSupp}
\end{figure}
\textit{Extraction of thresholds.} In the \textit{Hysteresis Measurement}, we extract the threshold for the onset of an imbalance \imb{} during ramp~I and the threshold where \imb{} vanishes again during ramp~II. We define both of these thresholds as the point where the intracavity photon signal is $20$ times higher than the mean background level. The background level is obtained by averaging the photon signal over $50$~ms while all lattices are switched off. As a result of this method, the imbalances \imb{} at the threshold positions are of different magnitude (see orange and green diamonds in Fig.~\ref{fig:hysteresis}A). The experiment is repeated at least $3$ times for every lattice depth \Vz{}, and the corresponding averaged thresholds for the $z$ lattice depth \Vz{} and detuning \Dc{} are shown by orange and green diamonds in Fig.~\ref{fig:phase_diagram}.

\textit{Hysteresis loop definition.} We show closed hysteresis loops of the imbalance \imb{} as a function of global-range interaction strength \Ul{} in Fig.~\ref{fig:hysteresis}A. The loop naturally closes at high \Ul{} (right side of the figure) where the detuning ramp is inverted. At low \Ul{} (left side of the figure) we plot data down to the point where the two curves cross. We only consider crossing points which happen below the thresholds of the creation and disappearance of an imbalance. This additional condition is needed to exclude crossings happening in the middle of the hysteresis loop due to e.g. heating, such a case is visible in Fig.~\ref{fig:hysteresis}D. In order to reduce noise, we average \Ul{} using an averaging window of $2\pi\times20$~Hz to find this crossing point.

\textit{Hysteresis area.} The hysteresis area $A$ is obtained by integrating the imbalance \imb{} as a function of \Ul{} during ramp~II and subtracting this signal from the integrated curve during ramp~I. We define a normalized hysteresis area as the ratio of hysteresis area $A$ and a factor $A_{\textsf{max}}$. Here $A_{\textsf{max}}$ is a fixed constant which defines the maximum possible hysteresis area, i.e. the product of the total number of atoms and the maximum strength of \Ul{}. The hysteresis area shown in Fig.~\ref{fig:hysteresis}B and Fig.~\ref{fig:timesSupp} is the average of at least $3$ repetitions for every lattice depth \Vz{}. In order to ensure comparability of the data, we use hysteresis area data only for those lattice depths for which the averaged maximum imbalance $\Theta_{\textsf{max}}$ satisfies the constraint that $\Theta_{\textsf{max}} \geq \Theta_{\textsf{z,max}} - \Delta \Theta_{\textsf{z,max}}$, where $\Theta_{\textsf{z,max}}$ is the average of the maximum imbalance obtained for the case of deepest lattices and $\Delta \Theta_{\textsf{z,max}}$ is the corresponding standard deviation.

At small \Ul{}, changes in the interaction strength stem from ramping the lattice depth \Vz{} which changes both \Ul{} and \Us{}. However, a large fraction of the hysteresis loop is occurring during the frequency ramps where \Ul{} is varying while \Us{} stays constant. Taking for example the case of a frequency ramp of duration $\tau=80$~ms as shown in Fig.~\ref{fig:hysteresis}, \Us{} is reduced by less then $9\%$ at the point where the hysteresis loop closes for small \Ul{}. At ramp times of $\tau=(30-50)$~ms, this reduction in \Us{} increases to $23\%$.

We note that we do not use \UlUs{} as an $x$ axis for the extraction of an hysteresis area as it does not allow a direct comparison between the case of strong short-range interactions \Us{} when all 3D lattices are present and the case where the $y$ lattice is switched off such as to reduce \Us{}. 

\textit{Hysteresis area as a function of ramp time $\tau$.} A study of the hysteresis area is shown in Fig.~\ref{fig:timesSupp}. In all cases, we observe a qualitatively comparable behavior as in Fig.~\ref{fig:hysteresis}B where increasing interactions increases the observed hysteresis area. Heating from the presence of optical lattices reduces the overall signal with increasing ramp time, leading to a negative hysteresis area clearly visible in Fig.~\ref{fig:timesSupp}C-D. The difference in the hysteresis area between the case of strong and weak \Us{} (with and without the $y$ lattice, respectively) is nearly the same for different ramp times. 

\section{Imbalance dynamics: data evaluation}
In order to quantify the position, duration and height of the fast jump as shown in Fig.~\ref{fig:imbalance_dynamics}, we use the following definition of an effective derivative:
\begin{align}
\frac{d \imb{}}{dt}\left(T\right)\Big|_\xi = &\frac{1}{\xi}\left[\mathrm{max}\big[ \imb{}(T-\frac{\xi}{2} : T+\frac{\xi}{2}) \big] 
- \mathrm{min}\big[ \imb{}(T-\frac{\xi}{2} : T+\frac{\xi}{2})  \big]\right]
\end{align}
where $\mathrm{max}$ and $\mathrm{min}$ yield the maximum and minimum value of \imb{} withing a time interval of $\pm\xi/2$ around the time $T$, we use $\xi=4$~ms. This effective derivative helps to improve the signal to noise ratio. The fast jump is then associated with a maximum in the amplitude of the effective derivative. We fit the signal from the effective derivative with a Gaussian in a time window of $\pm10$~ms around the fast jump. The central position of the Gaussian fit, $t_0$, is used to extract the position of the fast jump in terms of the $z$ lattice depth and the detuning \Dc{}, see Fig.~\ref{fig:phase_diagram}. The full width at half maximum of the Gaussian represents the duration of the jump $\xi_{\mathrm{jump}}$ and is used to extract the jump height 
\begin{equation}
h_{\mathrm{jump}}= \imb{}(T_0 + \xi_{\mathrm{jump}}/2) - \imb{}(T_0 - \xi_{\mathrm{jump}}/2)\nonumber
\end{equation}
In order to extract $T_0$, $\xi_{\mathrm{jump}}$, and $h_{\mathrm{jump}}$ we consider only those experimental realizations where the fast jump occurs when all the external parameters are kept constant after the quench, and we obtain $54$ such realizations. Formally, this constraint is defined as $T_0 - \xi_{\mathrm{jump}}/2 > T_{\mathrm{const}}$, where $T_{\mathrm{const}}$ is the time from which on all external parameters are kept constant. The height $\Delta\imb{}$ and duration $\Delta T$ of the fast jump stated in the main text are obtained by averaging all individual data of $h_{\mathrm{jump}}$ and $\xi_{\mathrm{jump}}$.

Repeating the extraction procedure described above with reduced time interval $\xi$ or reduced moving average window size, we observe shorter durations of the step at the cost of a reduced signal to noise ratio. The value provided in the main text is thus an upper bound on the actual step duration.\\

We obtain the tunneling time in a double well in the following way. We consider the two states $\left|1,1\right>$ and $\left|2,0\right>$ resonantly coupled by the tunneling $\sqrt{2} t$, where $\sqrt{2}$ accounts for bosonic enhancement. Starting in the state $\left|1,1\right>$, the system reaches the state $\left|2,0\right>$ within the tunneling time.\\

\textit{Phase of the light field:} Using our heterodyne detection we also extract the time phase of the light field scattered into the cavity with respect to the lattice in the $z$ direction \cite{Baumann2011a_sup}. Because of residual phase drifts of the heterodyne setup, we cannot relate the phase signals between consecutive experimental runs. To improve clarity of the phase signal shown in Fig.~\ref{fig:imbalance_dynamics}C a mean offset phase is subtracted in each realization to remove these shot to shot phase drifts. The mean offset phase is obtained by time averaging of the phase signal from $20\text{~ms} \leq T \leq 65\text{~ms}$ in the \textit{Metastability Measurement} and from $40\text{~ms} \leq T \leq 80\text{~ms}$ in the \textit{Hysteresis Measurement}. Here $T=0$~ms corresponds to the initial time (0~ms) in Fig.~\ref{fig:imbalance_dynamics}.

\section{Extraction of the change in excitation energy \dE{} from the measured photon flux}
We extract the change in excitation energy \dE{} during the imbalance jump (ii), where $\taus{}$ counts the time since the beginning of the jump. At the beginning of the jump, superfluid surface atoms account for an initial imbalance of $\imb{}(\taus{}=0\text{~ms})$. We assume that the imbalance stemming from these surface atoms stays approximately constant during the jump. As the imbalance stemming from reordering bulk atoms increases with time $\taus{}$ the site offset \del{off} also increases, reducing the excitation energy of all previously imbalanced atoms. From the measured imbalance $\imb{}(\taus{})$ and site offset $\del{off}(\taus{})$ we obtain
\begin{equation}
\frac{\dE{}(\taus{})}{h} = \underbrace{ \int\limits_{\taus{\prime}=0}^{\taus{}} \Big[ \del{off}(\taus{}) - \del{off}(\taus{\prime}) \Big] \frac{\text{d}\imb{}}{\text{d}\taus{\prime}} 
\text{d}\taus{\prime} }_{\text{bulk}}
+ \underbrace{ \vphantom{\int\limits_{\taus{\prime}=0}^{\taus{}}} \Big[ \del{off}(\taus{}) - \del{off}(0) \Big] \imb{}(0) }_{\text{surface}}
\label{eq:exc_energy}
\end{equation}
The result is shown in Fig.~\ref{fig:energy_dynamics}C.

\section{Number of photons scattered during the imbalance jump}
We estimate the total number of photons incoherently scattered from the $z$ lattice - off the atoms - into the single cavity mode. In the bad cavity limit we can consider a quasi-stationary intra-cavity light field. In this limit, scattering of photons into the cavity mode balances photon loss through the cavity mirrors. Photons leave the cavity at a rate given by the inverse cavity lifetime of $2\kappa=2\times2\pi\times 1.25$~MHz. Here we neglect the low rate of scattering of incoherent cavity photons back into the $z$ lattice as they will not exhibit bosonic enhancement. We observe an average mean intra-cavity photon number of $\nph{}=0.18(2)$ during the time of the imbalance jump (ii) of $\Delta T=4.3(0.6)$~ms. The scattering rate into the cavity then becomes $2.8(3) \times 10^6$~photons/s and the number of scattered photons during the jump is about $12(3) \times 10^3$~photons.

\section{Phase diagram measurement: data evaluation}
\begin{figure}[ht]
\centering
\includegraphics[width=87mm]{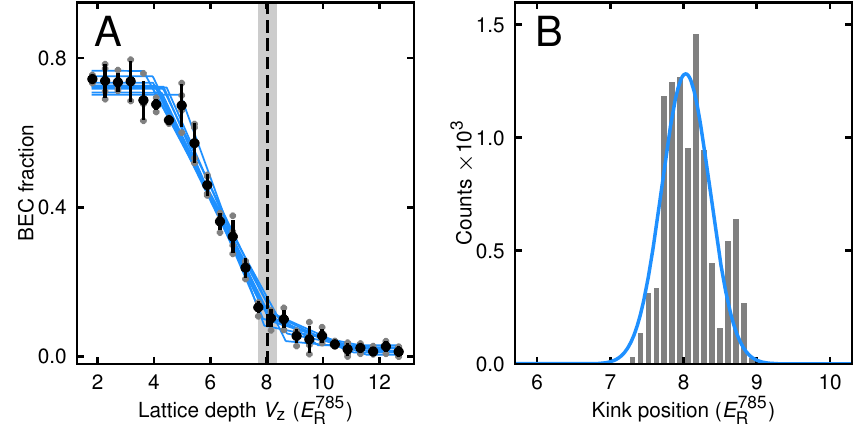}
\caption{Extraction of the phase boundary between states with and without spatial coherence. (A) The BEC fraction as a function of the $z$ lattice depth \Vz{} is shown for a detuning of $\Dctp{}=-13$~MHz. We observe a \textit{kink} in the BEC fraction and we use a multi-line fit to extract its position. The data is resampled $10^4$ times using a bootstrapping method and fitted separately to estimate the $1\sigma$ standard deviation (grey area) around the \textit{kink} position (dashed line). The blue lines represent the fit results of $10$ random samples. (B) Histogram of the \textit{kink} position resulting from resampling the data. We fit a Gaussian to the histogram and extract the \textit{kink} position and the $1\sigma$ standard deviation from this fit.} 
\label{fig:phase_boundary_spat_coh}
\end{figure}
To construct the phase diagram of the system we follow \cite{Landig2016_sup} with the difference that we now prepare a BEC of $16(1)\times 10^3$ atoms instead of $42(4)\times 10^3$ atoms. Due to the lower atom number, states with non-zero imbalance \imb{} are created closer to resonance with respect to \cite{Landig2016_sup}. The wavelength of the square lattice is now $784.7$~nm instead of $785.3$~nm previously while all other parameters are comparable.\\

Phase diagram measurement: Contrary to the \textit{Metastability Measurement} and the \textit{Hysteresis Measurement}, the detuning \Dc{} is kept constant throughout each experimental sequence. We start with a BEC and slowly ramp up the lattice depth in all three directions. Then, all trapping potentials are abrouptly switched off and absorption pictures of the atomic cloud are taken after $7$~ms of ballistic expansion. We obtain the BEC fraction from a bimodal fit of the atomic density distribution, and the maximum imbalance from the maximum photon flux leaking out of the cavity. To construct the phase diagram, this experiment is repeated at different detunings \Dc{} and final lattice depths \Vz{}.\\

Each data point of the phase diagram is taken on average four times. We obtain the phase boundary between states with and without spatial coherence for each detuning \Dc{} from the position of a \textit{kink} in the BEC fraction as a function of the lattice depth \Vz{}, see Fig.~\ref{fig:phase_boundary_spat_coh}, which we associate with the loss of superfluidity and the formation of an insulating phase \cite{Jimenez-Garcia2010_sup}. We use a multiple line fit to find the \textit{kink} position. For each detuning \Dc{} we estimate the standard deviation of the \textit{kink} position using a bootstrapping algorithm which resamples the data $10^4$ times. The samples are constructed by taking out of the four experimental iterations one data point in the BEC fraction at random for each lattice depth \Vz{}. The samples are then fitted individually and a histogram of the resulting \textit{kink} positions is constructed, see Fig.~\ref{fig:phase_boundary_spat_coh}. We obtain the position of the \textit{kink} and the $1~\sigma$ standard deviation shown in Fig.~\ref{fig:phase_diagram} from a Gaussian fit to the histogram.\\

We obtain information on the creation of an imbalance \imb{} by detecting photons leaking from the cavity with a heterodyne setup \cite{Landig2015_sup}. In each experimental repetition we take a single data point of \imb{} after all lattices are ramped up and just before taking atomic absorption pictures. The photon data is resampled together with the measured $z$ lattice depth \Vz{} in order to reduce noise. An averaging window of $10$~ms is used. We deduce the phase diagram from the imbalance \imb{} and the transition between states with and without spatial coherence, where we use the criteria established in \cite{Landig2016_sup}: the superfluid region (SF) shows spatial coherence but no imbalance, the lattice supersolid region (SS) shows spatial coherence and a non-zero imbalance, the Mott-insulating region (MI) shows no spatial coherence and no imbalance and the charge-density wave region (CDW) shows no spatial coherence but a non-zero imbalance.


\begin{thebibliography}{10}

\bibitem{Anderson1972}
Anderson, P.~w., Halperin, B.~I., and Varma, c.~M.
\newblock {\em Philosophical Magazine}{ \bf 25}(1), 1--9   (1972).

\bibitem{Karplus2002}
Karplus, M. and McCammon, J.~A.
\newblock {\em Nature Structural Biology}{ \bf 9}(9), 646--652   (2002).

\bibitem{Brazhkin2006}
Brazhkin, V.~V.
\newblock {\em Journal of Physics: Condensed Matter}{ \bf 18}(42), 9643--9650
    (2006).

\bibitem{Binder1987}
Binder, K.
\newblock {\em Reports on Progress in Physics}{ \bf 50}(7), 783--859 
  (1987).

\bibitem{Menotti2007}
Menotti, C., Trefzger, C., and Lewenstein, M.
\newblock {\em Physical Review Letters}{ \bf 98}(23), 235301   (2007).

\bibitem{Gopalakrishnan2011}
Gopalakrishnan, S., Lev, B.~L., and Goldbart, P.~M.
\newblock {\em Physical Review Letters}{ \bf 107}(27), 277201   (2011).

\bibitem{Strack2011}
Strack, P. and Sachdev, S.
\newblock {\em Physical Review Letters}{ \bf 107}(27), 277202   (2011).

\bibitem{Altman2015}
Altman, E. and Vosk, R.
\newblock {\em Annual Review of Condensed Matter Physics}{ \bf 6}(1), 383--409
    (2015).

\bibitem{Andraschko2014}
Andraschko, F., Enss, T., and Sirker, J.
\newblock {\em Physical Review Letters}{ \bf 113}(21), 217201   (2014).

\bibitem{Eisert2015}
Eisert, J., Friesdorf, M., and Gogolin, C.
\newblock {\em Nature Physics}{ \bf 11}(2), 124--130   (2015).

\bibitem{Haller2009}
Haller, E., Gustavsson, M., Mark, M.~J., Danzl, J.~G., Hart, R., Pupillo, G.,
  and Nagerl, H.-C.
\newblock {\em Science}{ \bf 325}(5945), 1224--1227   (2009).

\bibitem{Eckel2014}
Eckel, S., Lee, J.~G., Jendrzejewski, F., Murray, N., Clark, C.~W., Lobb,
  C.~J., Phillips, W.~D., Edwards, M., and Campbell, G.~K.
\newblock {\em Nature}{ \bf 506}(7487), 200--203   (2014).

\bibitem{Schreiber2015}
Schreiber, M., Hodgman, S.~S., Bordia, P., Luschen, H.~P., Fischer, M.~H.,
  Vosk, R., Altman, E., Schneider, U., and Bloch, I.
\newblock {\em Science}{ \bf 349}(6250), 842--845   (2015).

\bibitem{Campbell2016}
Campbell, D.~L., Price, R.~M., Putra, A., Vald{\'{e}}s-Curiel, A.,
  Trypogeorgos, D., and Spielman, I.~B.
\newblock {\em Nature Communications}{ \bf 7}, 10897   (2016).

\bibitem{Kadau2016}
Kadau, H., Schmitt, M., Wenzel, M., Wink, C., Maier, T., Ferrier-Barbut, I.,
  and Pfau, T.
\newblock {\em Nature}{ \bf 530}(7589), 194--197   (2016).

\bibitem{Trenkwalder2016}
Trenkwalder, A., Spagnolli, G., Semeghini, G., Coop, S., Landini, M., Castilho,
  P., Pezz{\`{e}}, L., Modugno, G., Inguscio, M., Smerzi, A., and Fattori, M.
\newblock {\em Nature Physics}{ \bf 12}(9), 826--829   (2016).

\bibitem{Letscher2016}
Letscher, F., Thomas, O., Niederpr{\"{u}}m, T., Fleischhauer, M., and Ott, H.
\newblock {\em Physical Review X}{ \bf 7}(2), 021020   (2017).

\bibitem{Antoni1995}
Antoni, M. and Ruffo, S.
\newblock {\em Physical Review E}{ \bf 52}(3), 2361--2374   (1995).

\bibitem{Mukamel2005}
Mukamel, D., Ruffo, S., and Schreiber, N.
\newblock {\em Physical Review Letters}{ \bf 95}(24), 240604   (2005).

\bibitem{Baumann2010}
Baumann, K., Guerlin, C., Brennecke, F., and Esslinger, T.
\newblock {\em Nature}{ \bf 464}(7293), 1301--1306   (2010).

\bibitem{Mottl2012}
Mottl, R., Brennecke, F., Baumann, K., Landig, R., Donner, T., and Esslinger,
  T.
\newblock {\em Science}{ \bf 336}(6088), 1570--1573   (2012).

\bibitem{Klinder2015insulator}
Klinder, J., Ke{\ss}ler, H., Bakhtiari, M.~R., Thorwart, M., and Hemmerich, A.
\newblock {\em Physical Review Letters}{ \bf 115}(23), 230403   (2015).

\bibitem{Landig2016}
Landig, R., Hruby, L., Dogra, N., Landini, M., Mottl, R., Donner, T., and
  Esslinger, T.
\newblock {\em Nature}{ \bf 532}(7600), 476--479   (2016).

\bibitem{Li2013}
Li, Y., He, L., and Hofstetter, W.
\newblock {\em Physical Review A}{ \bf 87}(5), 051604   (2013).

\bibitem{Bakhtiari2015}
Bakhtiari, M.~R., Hemmerich, A., Ritsch, H., and Thorwart, M.
\newblock {\em Physical Review Letters}{ \bf 114}(12), 123601   (2015).

\bibitem{Caballero-Benitez2015}
Caballero-Benitez, S.~F. and Mekhov, I.~B.
\newblock {\em Physical Review Letters}{ \bf 115}(24), 243604   (2015).

\bibitem{Chen2016}
Chen, Y., Yu, Z., and Zhai, H.
\newblock {\em Physical Review A}{ \bf 93}(4), 041601   (2016).

\bibitem{Dogra2016}
Dogra, N., Brennecke, F., Huber, S.~D., and Donner, T.
\newblock {\em Physical Review A}{ \bf 94}(2), 023632   (2016).

\bibitem{Niederle2016}
Niederle, A.~E., Morigi, G., and Rieger, H.
\newblock {\em Physical Review A}{ \bf 94}(3), 033607   (2016).

\bibitem{Sundar2016}
Sundar, B. and Mueller, E.~J.
\newblock {\em Physical Review A}{ \bf 94}(3), 033631   (2016).

\bibitem{Panas2017}
Panas, J., Kauch, A., and Byczuk, K.
\newblock {\em Physical Review B}{ \bf 95}(11), 115105   (2017).

\bibitem{Flottat2017}
Flottat, T., de~Parny, L. d.~F., H{\'{e}}bert, F., Rousseau, V.~G., and
  Batrouni, G.~G.
\newblock {\em Physical Review B}{ \bf 95}(14), 144501   (2017).

\bibitem{Jaksch1998}
Jaksch, D., Bruder, C., Cirac, J.~I., Gardiner, C.~W., and Zoller, P.
\newblock {\em Physical Review Letters}{ \bf 81}(15), 3108--3111   (1998).

\bibitem{Greiner2002}
Greiner, M., Mandel, O., Esslinger, T., H{\"{a}}nsch, T.~W., and Bloch, I.
\newblock {\em Nature}{ \bf 415}(6867), 39--44   (2002).

\bibitem{Landig2015}
Landig, R., Brennecke, F., Mottl, R., Donner, T., and Esslinger, T.
\newblock {\em Nature Communications}{ \bf 6}, 7046   (2015).

\bibitem{Klinder2015}
Klinder, J., Ke{\ss}ler, H., Wolke, M., Mathey, L., and Hemmerich, A.
\newblock {\em Proceedings of the National Academy of Sciences}{ \bf 112}(11),
  3290--3295   (2015).

\bibitem{Lipowsky1983}
Lipowsky, R. and Speth, W.
\newblock {\em Physical Review B}{ \bf 28}(7), 3983--3993   (1983).

\bibitem{Lipowsky1987}
Lipowsky, R.
\newblock {\em Ferroelectrics}{ \bf 73}(1), 69--81   (1987).

\bibitem{Hung2010}
Hung, C.-L., Zhang, X., Gemelke, N., and Chin, C.
\newblock {\em Physical Review Letters}{ \bf 104}(16), 160403   (2010).

\bibitem{Baumann2011a}
Baumann, K., Mottl, R., Brennecke, F., and Esslinger, T.
\newblock {\em Physical Review Letters}{ \bf 107}(14), 140402   (2011).

\end{thebibliography}

\begin{thebibliography}{1}

\bibitem{Stoferle2004_sup}
St{\"{o}}ferle, T., Moritz, H., Schori, C., K{\"{o}}hl, M., and Esslinger, T.
\newblock {\em Physical Review Letters}{ \bf 92}(13), 130403   (2004).

\bibitem{Morsch2006_sup}
Morsch, O. and Oberthaler, M.
\newblock {\em Reviews of Modern Physics}{ \bf 78}(1), 179--215   (2006).

\bibitem{Landig2016_sup}
Landig, R., Hruby, L., Dogra, N., Landini, M., Mottl, R., Donner, T., and
  Esslinger, T.
\newblock {\em Nature}{ \bf 532}(7600), 476--479   (2016).

\bibitem{Dogra2016_sup}
Dogra, N., Brennecke, F., Huber, S.~D., and Donner, T.
\newblock {\em Physical Review A}{ \bf 94}(2), 023632   (2016).

\bibitem{Panas2017_sup}
Panas, J., Kauch, A., and Byczuk, K.
\newblock {\em Physical Review B}{ \bf 95}(11), 115105   (2017).

\bibitem{Pedri2001_sup}
Pedri, P., Pitaevskii, L., Stringari, S., Fort, C., Burger, S., Cataliotti,
  F.~S., Maddaloni, P., Minardi, F., and Inguscio, M.
\newblock {\em Physical Review Letters}{ \bf 87}(22), 220401   (2001).

\bibitem{Dhar2011_sup}
Dhar, A., Singh, M., Pai, R.~V., and Das, B.~P.
\newblock {\em Physical Review A}{ \bf 84}(3), 033631   (2011).

\bibitem{Baumann2011a_sup}
Baumann, K., Mottl, R., Brennecke, F., and Esslinger, T.
\newblock {\em Physical Review Letters}{ \bf 107}(14), 140402   (2011).

\bibitem{Jimenez-Garcia2010_sup}
Jim{\'{e}}nez-Garc{\'{i}}a, K., Compton, R.~L., Lin, Y.-J., Phillips, W.~D.,
  Porto, J.~V., and Spielman, I.~B.
\newblock {\em Physical Review Letters}{ \bf 105}(11), 110401   (2010).

\bibitem{Landig2015_sup}
Landig, R., Brennecke, F., Mottl, R., Donner, T., and Esslinger, T.
\newblock {\em Nature Communications}{ \bf 6}, 7046   (2015).

\end{thebibliography}
\end{document}